\documentclass[longauth]{aa}

 \usepackage{ gensymb }
 \usepackage{graphicx}
 \usepackage{natbib,twoopt}
 \usepackage{amsmath}
 \usepackage{ textcomp }
 \usepackage{pgf}
 \usepackage[breaklinks=true]{hyperref} 
 \bibpunct{(}{)}{;}{a}{}{,}    
 \newcommandtwoopt{\citeads}[3][][]{\href{http://adsabs.harvard.edu/abs/#3}%
                                        {\citealp[#1][#2]{#3}}}
 \newcommandtwoopt{\citepads}[3][][]{\href{http://adsabs.harvard.edu/abs/#3}%
                                        {\citep[#1][#2]{#3}}}
 \newcommandtwoopt{\citetads}[3][][]{\href{http://adsabs.harvard.edu/abs/#3}%
                                        {\citet[#1][#2]{#3}}}
 \newcommandtwoopt{\citeyearads}[3][][]%
   {\href{http://adsabs.harvard.edu/abs/#3}{\citeyear[#1][#2]{#3}}}

\usepackage{amstext}

\begin{document}

\titlerunning{Shock location and CME 3D reconstruction of a solar type II radio burst with LOFAR.}
\authorrunning{Zucca et al.}
\title{Shock location and CME 3D reconstruction of a solar type II radio burst with LOFAR.}

\author{P.~Zucca\inst{1}
\and D. E.~Morosan\inst{2}
\and A.~P.~Rouillard\inst{3} 
\and R.~Fallows\inst{1}
\and P.~T. Gallagher\inst{2}
\and J.~Magdalenic\inst{4}
\and K-L.~Klein\inst{5}
\and G.~Mann\inst{6}
\and C.~Vocks\inst{6}
\and E.~P. Carley\inst{2}
\and M. M.~Bisi\inst{7}
\and E.~P. Kontar\inst{8}
\and H.~Rothkaehl\inst{9}
\and B.~Dabrowski\inst{10}
\and A.~Krankowski\inst{10}
\and J.~Anderson\inst{11}
\and A.~Asgekar\inst{1,12}
\and M.~E.~Bell\inst{13}
\and M.~J.~Bentum\inst{1,14}
\and P.~Best\inst{15}
\and R.~Blaauw\inst{1}
\and F.~Breitling\inst{6}
\and J.~W.~Broderick\inst{1}
\and W.~N.~Brouw\inst{1,16}
\and M.~Br\"uggen\inst{17}
\and H.~R.~Butcher\inst{18}
\and B.~Ciardi\inst{19}
\and E.~de Geus\inst{1,20}
\and A.~Deller\inst{21,1}
\and S.~Duscha\inst{1}
\and J.~Eisl\"offel\inst{22}
\and M.~A.~Garrett\inst{23,24}
\and J.~M.~Grie\ss{}meier\inst{25,26}
\and A.~W.~Gunst\inst{1}
\and G.~Heald\inst{27,1}
\and M.~Hoeft\inst{14}
\and J.~H\"orandel\inst{28}
\and M.~Iacobelli\inst{1}
\and E.~Juette\inst{29}
\and A. ~Karastergiou\inst{30}
\and J.~van Leeuwen\inst{2,31}
\and D.~McKay-Bukowski\inst{32,33}
\and H. Mulder\inst{1}
\and H.~Munk\inst{34,1}
\and A.~Nelles\inst{35}
\and E.~Orru\inst{1}
\and H.~Paas\inst{36}
\and V.~N.~Pandey\inst{1,8}
\and R.~Pekal\inst{37}
\and R.~Pizzo\inst{1}
\and A.~G.~Polatidis\inst{1}
\and W.~Reich\inst{38}
\and A.~ Rowlinson\inst{1}
\and D.~J.~Schwarz\inst{39}
\and A.~Shulevski\inst{16}
\and J.~Sluman\inst{1}
\and O.~Smirnov\inst{40,41}
\and C.~Sobey\inst{42}
\and M.~Soida\inst{43}
\and S.~Thoudam\inst{44}
\and M.~C.~Toribio\inst{16,1}
\and R.~Vermeulen\inst{1}
\and R.~J.~van Weeren\inst{16}
\and O.~Wucknitz\inst{38}
\and P.~Zarka\inst{5}
}

\institute {ASTRON, Netherlands Institute for Radio Astronomy, Postbus 2, 7990 AA, Dwingeloo, The Netherlands.
\and Astrophysics Research Group, School of Physics, Trinity College Dublin, Dublin 2, Ireland.
\and Institut de Recherche en Astrophysique et Planetologie, 9 Ave. du Colonel Roche 31028, Toulouse Cedex 4, France.
\and Solar-Terrestrial Center of Excellence, Royal Observatory of Belgium, Avenue Circulaire 3, 1180 Brussels, Belgium.
\and LESIA, UMR CNRS 8109, Observatoire de Paris, 92195 Meudon, France.
\and Leibniz-Institut f\"ur Astrophysik Potsdam (AIP), An der Sternwarte 16, 14482 Potsdam, Germany.
\and RAL Space, Science and Technology Facilities Council, Rutherford Appleton Laboratory, Oxfordshire, UK.
\and SUPA School of Physics and Astronomy, University of Glasgow, G12 8QQ, United Kingdom.
\and Space research Centre of the Polish Academy of Science, 18A Bartycka 00-716 Warsaw, Poland. 
\and Space Radio-Diagnostics Research Centre, University of Warmia and Mazury in Olsztyn, Poland. 
\and Helmholtz-Zentrum Potsdam, DeutschesGeoForschungsZentrum GFZ, Geodesy and Remote Sensing, Telegrafenberg, A17, 14473 Potsdam, Germany.
\and Shell Technology Center, Bangalore, India.
\and University of Technology Sydney, 15 Broadway, Ultimo NSW 2007, Australia.
\and Eindhoven University of Technology, P.O. Box 513, 5600 MB  Eindhoven, The Netherlands.
\and Institute for Astronomy, University of Edinburgh, Royal Observatory of Edinburgh, Blackford Hill, Edinburgh EH9 3HJ, UK.
\and Kapteyn Astronomical Institute, PO Box 800, 9700 AV Groningen, The Netherlands. 
\and University of Hamburg, Gojenbergsweg 112, 21029 Hamburg, Germany. 
\and Research School of Astronomy and Astrophysics, Australian National University, Canberra, ACT 2611 Australia. 
\and Max Planck Institute for Astrophysics, Karl Schwarzschild Str. 1, 85741 Garching, Germany. 
\and SmarterVision BV, Oostersingel 5, 9401 JX Assen. 
\and Centre for Astrophysics \& Supercomputing, Swinburne University of Technology John St, Hawthorn VIC 3122 Australia. 
\and Th\"{u}ringer Landessternwarte, Sternwarte 5, D-07778 Tautenburg, Germany. 
\and Jodrell Bank Center for Astrophysics, School of Physics and Astronomy, The University of Manchester, Manchester M13 9PL,UK. 
\and Leiden Observatory, Leiden University, PO Box 9513, 2300 RA Leiden, The Netherlands. 
\and LPC2E - Universite d'Orleans/CNRS. 
\and Station de Radioastronomie de Nancay, Observatoire de Paris - CNRS/INSU, USR 704 - Univ. Orleans, OSUC , route de Souesmes, 18330 Nancay, France. 
\and CSIRO Astronomy and Space Science, 26 Dick Perry Avenue, Kensington, WA 6151, Australia.  
\and Department of Astrophysics/IMAPP, Radboud University Nijmegen, P.O. Box 9010, 6500 GL Nijmegen, The Netherlands. 
\and Astronomisches Institut der Ruhr-Universit\"{a}t Bochum, Universitaetsstrasse 150, 44780 Bochum, Germany. 
\and Astrophysics, University of Oxford, Denys Wilkinson Building, Keble Road, Oxford OX1 3RH. 
\and Anton Pannekoek Institute for Astronomy, University of Amsterdam, Science Park 904, 1098 XH Amsterdam, The Netherlands. 
\and Department of Physics and Technology, University of Tromsø, Norway.
\and STFC Rutherford Appleton Laboratory,  Harwell Science and Innovation Campus,  Didcot  OX11 0QX, UK. 
\and Radboud University Radio Lab, Nijmegen, P.O. Box 9010, 6500 GL Nijmegen, The Netherlands. 
\and Department of Physics and Astronomy, University of California Irvine, Irvine, CA 92697, USA. 
\and Center for Information Technology (CIT), University of Groningen, The Netherlands. 
\and Poznan Supercomputing and Networking Center (PCSS) Poznan, Poland. 
\and Max-Planck-Institut f\"{u}r Radioastronomie, Auf dem H\"ugel 69, 53121 Bonn, Germany. 
\and Fakult\"{a}t f\''{u}r Physik, Universit\"{a}t Bielefeld, Postfach 100131, D-33501, Bielefeld, Germany. 
\and Department of Physics and Elelctronics, Rhodes University, PO Box 94, Grahamstown 6140, South Africa. 
\and SKA South Africa, 3rd Floor, The Park, Park Road, Pinelands, 7405, South Africa. 
\and International Centre for Radio Astronomy Research - Curtin University, GPO Box U1987, Perth, WA 6845, Australia. 
\and Jagiellonian University, Astronomical Observatory, Orla 171, PL 30-244 Krakow, Poland. 
\and Department of Physics and Electrical Engineering, Linnaeus University 35195, Vaexjoe, Sweden.
}


\abstract{Type II radio bursts are evidence of shocks in the solar atmosphere and inner heliosphere that emit radio waves ranging from sub-meter to kilometer lengths. These shocks may be associated with coronal mass ejections (CMEs) and reach speeds higher than the local magnetosonic speed. Radio imaging of decameter wavelengths (20--90~MHz) is now possible with the Low Frequency Array (LOFAR), opening a new radio window in which to study coronal shocks that leave the inner solar corona and enter the interplanetary medium and to understand their association with CMEs.}{To this end, we study a coronal shock associated with a CME and type II radio burst to determine the locations at which the radio emission is generated, and we investigate the origin of the band-splitting phenomenon.}{The type II shock source-positions and spectra were obtained using 91 simultaneous tied-array beams of LOFAR, and the CME was observed by the Large Angle and Spectrometric Coronagraph (LASCO) on board the \emph{Solar and Heliospheric Observatory} ($SOHO$) and by the COR2A coronagraph of the SECCHI instruments on board the \emph{Solar Terrestrial Relation Observatory} ($STEREO$). The 3D structure was inferred using triangulation of the coronographic observations. Coronal magnetic fields were obtained from a 3D magnetohydrodynamics (MHD) polytropic model using the photospheric fields measured by the Heliospheric Imager ($HMI$) on board the \emph{Solar Dynamic Observatory} ($SDO$) as lower boundary.}{The type II radio source of the coronal shock observed between 50 and 70~MHz was found to be located at the expanding flank of the CME, where  the shock geometry is quasi-perpendicular with $\theta_{\rm Bn} \sim 70^\circ$. The type II radio burst showed first and second harmonic emission; the second harmonic source was cospatial with the first harmonic source to within the observational uncertainty. This suggests that radio wave propagation does not alter the apparent location of the harmonic source. The sources of the two split bands were also found to be cospatial within the observational uncertainty, in agreement with the interpretation that split bands are simultaneous radio emission from upstream and downstream of the shock front. The fast magnetosonic Mach number derived from this interpretation was found to lie in the range 1.3-1.5. The fast magnetosonic Mach numbers derived from modelling the CME and the coronal magnetic field around the type II source were found to lie in the range 1.4-1.6.}{}

\maketitle

Accepted 26 March 2018

\section{Introduction}
Type II radio bursts are the result of magnetohydrodynamics  (MHD) shocks in the solar atmosphere \citep{Uchida1960,Wild1962,Mann1995}, and they can be observed to range from sub-metric to hectometric  wavelengths ($\sim$400~MHz to $\sim$0.4~MHz). Several candidates for triggering and driving these MHD shocks have been proposed, such as coronal mass ejections (CMEs), flares, coronal waves,  erupting loops or plasmoids, ejecta-like sprays, and X-ray jets  \citep[][and references therein]{Pick2008,Nindos2008}. \citet{Magdalenic2010} showed that in addition to type II bursts related to CMEs, flares may also be responsible for the production of a shock wave that drives the type II bursts. However, several statistical studies \citep{Classen2002, Cho2005, Gopalswamy2006_2} and recent case studies \citep{Zimovets2012,Zucca2014b,Pick2016} using radio spectral observations together with white-light and X-ray images showed that CMEs can initiate most of the metric type II (m-type II) bursts.

The region of the CME that is responsible for driving the shock might be different for each event and has not yet been comprehensively identified. Multiple scenarios have been suggested, such as a  pure bow shock at the CME front and a multiple shock scenario in internal parts, or flanks of the CME, possibly also related to blast waves as triggering events \citep[see e.g.][]{Classen2002,Nindos2011}.  Furthermore, as the shock is triggered by a propagating front travelling faster than the magnetosonic wave speed, travelling disturbances in the corona can create shocks only in specific structures with low Alfv\'en speed. In some cases, the propagating wave can also steepen into a shock when it moves towards an environment with decreasing Alv\'en speed \citep[see e.g.][]{Vrsnak_Lulic2000}. The scenario is then additionally complicated as the electron acceleration at shocks that results in radio emission may be restricted to quasi-perpendicular regions \citep{Holman1983,Bale1999}. Therefore, a detailed analysis by interferometric radio observations, extreme ultraviolet (EUV) and white-light together with models or data-constrained models of the magnetic field and Alfv\'en speed are necessary to fully understand the type II burst -- CME paradigm. Several cases of m-type II bursts have been studied using radio positional information at frequencies $\ge$ 150~MHz with the Nanc\c ay radio-heliograph \citep[NRH;][]{Kerdraon1997}. In these studies, the radio source location is compared with EUV or X--ray observations \citep[e.g. ][]{Klein1999,Dauphin2006,Nindos2011, Zimovets2012, Zucca2014b}, while only a few cases of radio burst imaging and white-light CMEs are available \citep[e.g.][]{Maia2000}. The lack of radio positional information compared with white-light CMEs is mainly due to observational constraints. Metric type II burst typically occur in the low corona (i.e. $<2$~$R_{\odot}$); these heights are currently occulted in space-borne coronagraphs, while for type II in the deca-hectometric range (DH-type II) at heights $>2$~$R_{\odot}$, where coronagraphs are available, radio imaging observations are unavailable. Radio-heliographic observations of type II radio bursts at 109~MHz have been reported by \citet{Ramesh2012} with the Gauribidanur radio-heliograph \citep{Ramesh1998}, and there are a few observations in the literature of type II radio bursts at 80~MHz \citep[e.g.][]{Gary1984} using the no longer operating Culgoora radio heliograph \citep{Wild1967}.  To date, there is no radio positional observation of type II bursts at frequencies $<80$~MHz. 
A range of low-frequency radio imaging arrays have been developed in the past years, such as the Murchison Widefield Array \citep[MWA;][]{Tingay2013}, which was recently used for solar observations \citep[e.g.][]{Mohan2017}, and the LOw Frequency ARray \citep[LOFAR;][]{vanHaarlem2013}. LOFAR operates at frequencies of 10--240 MHz, and it features multi-beaming capabilities, which can be used to produce heliographic imaging of the radio source \citep{Morosan2014,Morosan2015,Reid2017}.  The frequency domain at which LOFAR operates bridges the gap between the metric band and the currently unexplored imaging of the decametric band.

We here use LOFAR tied-array beam imaging and spectroscopy to study the location of a decametric type II radio burst and understand the region of the CME responsible for triggering the shock and the role of the ambient magnetic field and fast magnetosonic speed. In Section 2 we give an overview of the observational method, and we present the results of the tied-array beam imaging analysis and of the 3D reconstruction of the CME. In Section 3 we discuss the results and present the conclusion. 

\begin{figure}
\begin{center}
\includegraphics[trim=1cm 4.5cm 1cm 1cm,clip=true,width=8.5cm,angle=-90]{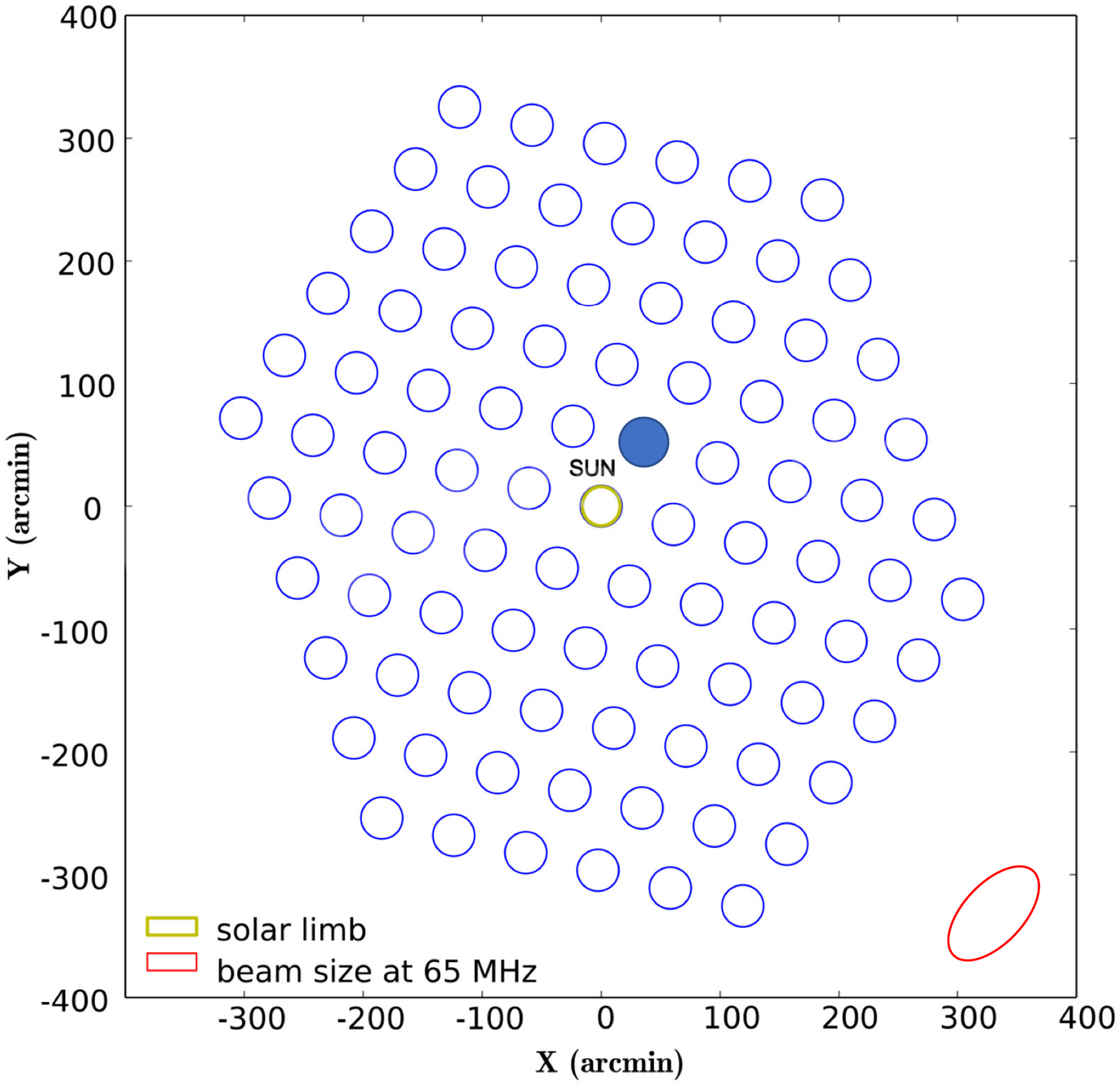}
\caption{Map of 91 tied-array beams covering a field of view of $\sim$16~$R_{\odot}$ centred on the Sun. The FWHM of the beam at a frequency of 65 MHz is represented by the red circle, and the size of the optical Sun is represented by the yellow circle. The blue filled dot is the location of the beam at which the dynamic spectrum of Figure~\ref{spectra} was obtained.}
\label{beam_pattern}
\end{center}
\end{figure}

\section{Observations and data analysis}

\label{observations}

\subsection{LOFAR observations}
On 2013 October 26, a type II radio bursts was recorded using one of the LOFAR beam-formed modes \citep{Stappers2011,vanHaarlem2013}. The radio burst was observed with the Low Band Antennas (LBAs) operating at frequencies of 10--90 MHz using six stations at the heart of the core combined to effectively form a single large station, a 320 m diameter island referred to as the Superterp.

\begin{figure}
\begin{center}
\includegraphics[trim=3cm 4.5cm 3cm 4cm,clip=true,width=6.5cm,angle=-90]{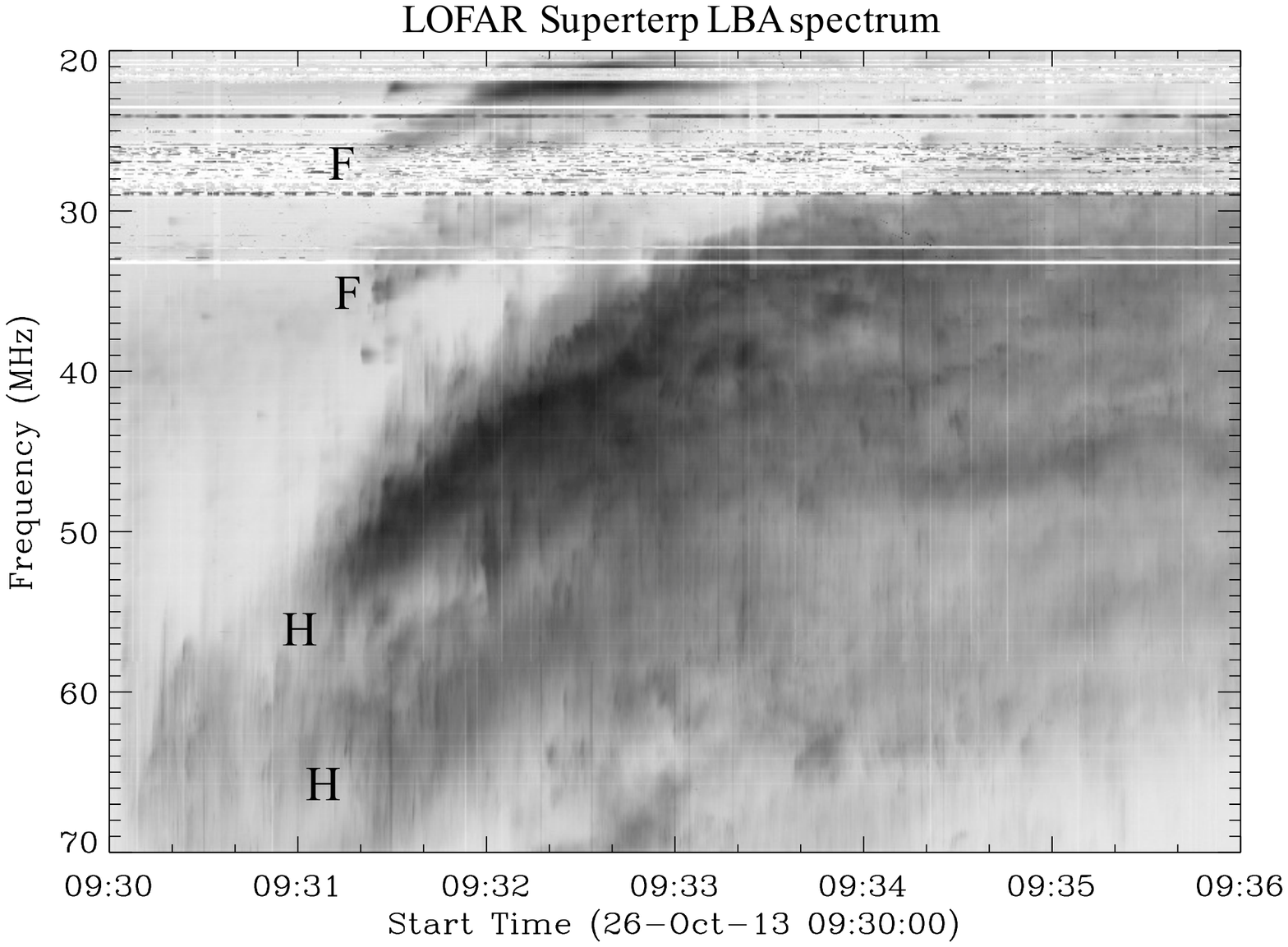}
\caption{Dynamic spectrum (from the beam reported with the filled blue circle in Figure~\ref{beam_pattern}) of a type II radio burst recorded on 2013 October 26 at 9:30 UT, showing fundamental (F) and harmonic (H) components, both split into two lanes. The spectrum was obtained with the LOFAR Superterp LBA antennas. The decrease in sensitivity below 29 MHz is due to the filter for the HF band.}
\label{spectra}
\end{center}
\end{figure}

We used 91 simultaneous beams to cover a field of view of $\sim$16~$R_{\odot}$ centred on the Sun. Each beam produces a dynamic spectrum with high time- and frequency resolution (10 ms; 12.5 kHz) at a unique spatial location that can be used to produce tied-array images of radio bursts \citep[see ][]{Morosan2014,Morosan2015}. 

\begin{figure*}[t]
\begin{center}
\includegraphics[trim=3.5cm 0cm 3.5cm 0cm,clip=true,width=8.5cm,angle=-90]{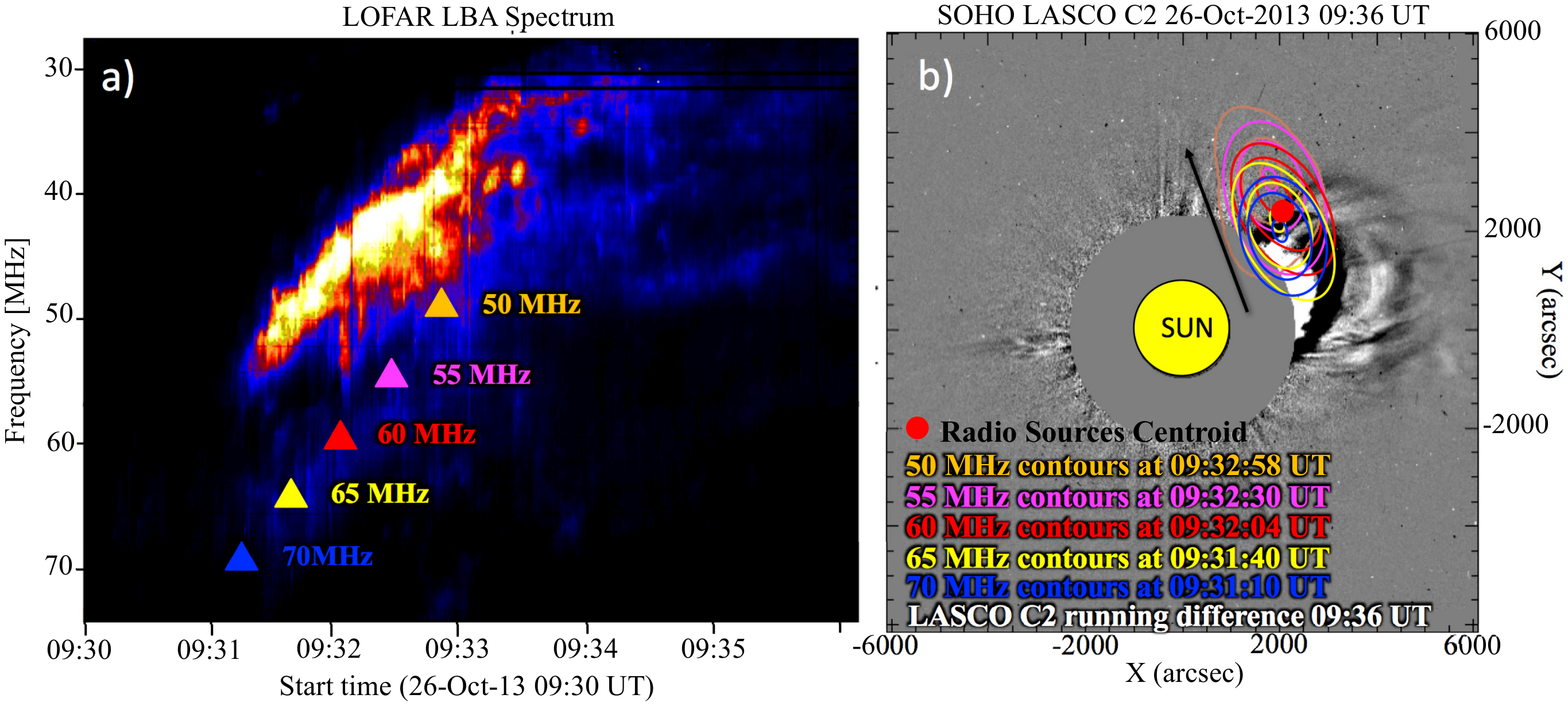}
\caption{(a) Dynamic spectrum of the type II radio burst observed on 2013 Oct 26; the harmonic emission is visible from 70 to $\sim$30~MHz. The specific times and frequencies at which the location of the radio source is calculated are indicated with coloured triangles. (b) Running-difference image of the CME observed with $SOHO$/LASCO (09:36--09:24 UT) with superposed contours of the radio sources (80\%, 90\%, and 95\%) using the same colour code as for the triangles in panel (a).}
\label{spectrum_location}
\end{center}
\end{figure*}

The FWHM of the tied-array beam size with this beam configuration is estimated to lie between 1.1 and 2.2~$R_{\odot}$ from 80 to 50 MHz because of the reduced spatial resolution of the tied-array beam imaging, which  uses a baseline of $\sim$320 m. The location of the 91 beams is shown in Figure~\ref{beam_pattern}. The dynamic spectrum of the type II radio burst is shown in Figure~\ref{spectra}. The burst shows fundamental (F) and harmonic (H) emission, and both lanes present band-splitting. Radio emission started around 9:30 UT, and the harmonic component drifted from 70 MHz to 30 MHz in approximately 6 minutes. 
The intensity of the radiation at a specific beam location can then be used to produce a ``macro-pixel'' map at a chosen time and frequency.  Figure~\ref{spectrum_location}a shows the dynamic spectrum of the type II burst with the superposed triangles indicating the frequency and time at which the locations of the type II burst are reported in Figure~\ref{spectrum_location}b.

The location of the radio source for this specific setting of beam locations (using only the Superterp) can be estimated to
lie between 70 and 50~MHz, while below 50~MHz, the larger source size results in emission spilling in the adjacent beam side-lobes.  The location of sources below 50~MHz with the tied-array beam mode configuration requires a larger baseline than the Superterp (see \citet{Morosan2014,Morosan2015} for full core tied-array observations) or a full knowledge of the beam shape. Figure~\ref{spectrum_location}b shows the contours (80\%, 90\%, and 95\%) of the radio emission flux at 70, 65, 60, 55, and 50~MHz, following the harmonic emission band from 09:31:10 UT to 09:32:58 UT. These contours are superposed to the running difference image of the CME observed with $SOHO$/LASCO (09:36 - 09:24 UT). The radio sources are located in the flank of the CME and appear to show a trend of motion consistent with the lateral expansion of the flank (indicated by the black arrow). However, with this set of observations and with the effects of scattering and refraction in the corona at these wavelengths (\citet{Kontar2017} report a size increase around 20 arcminutes and a potential shift up to $\sim$5 arcminutes for LOFAR frequencies), we do not take into account any source motion for this study. When metric or decametric radiation propagates through the corona, it is both refracted and scattered by turbulent plasma processes that may affect the apparent positions of the radio sources. For this reason, we did not estimate the speed of the radio sources but we compared their positional centroid (indicated in Figure~\ref{spectrum_location}b by the red dot) with white-light observations. The position of the type II radio burst at different frequencies was found to be located within the uncertainties at the flank of the CME.

\begin{figure}
\begin{center}
\includegraphics[trim=2.2cm 7cm 2cm 7cm,clip=true,width=8cm,angle=0]{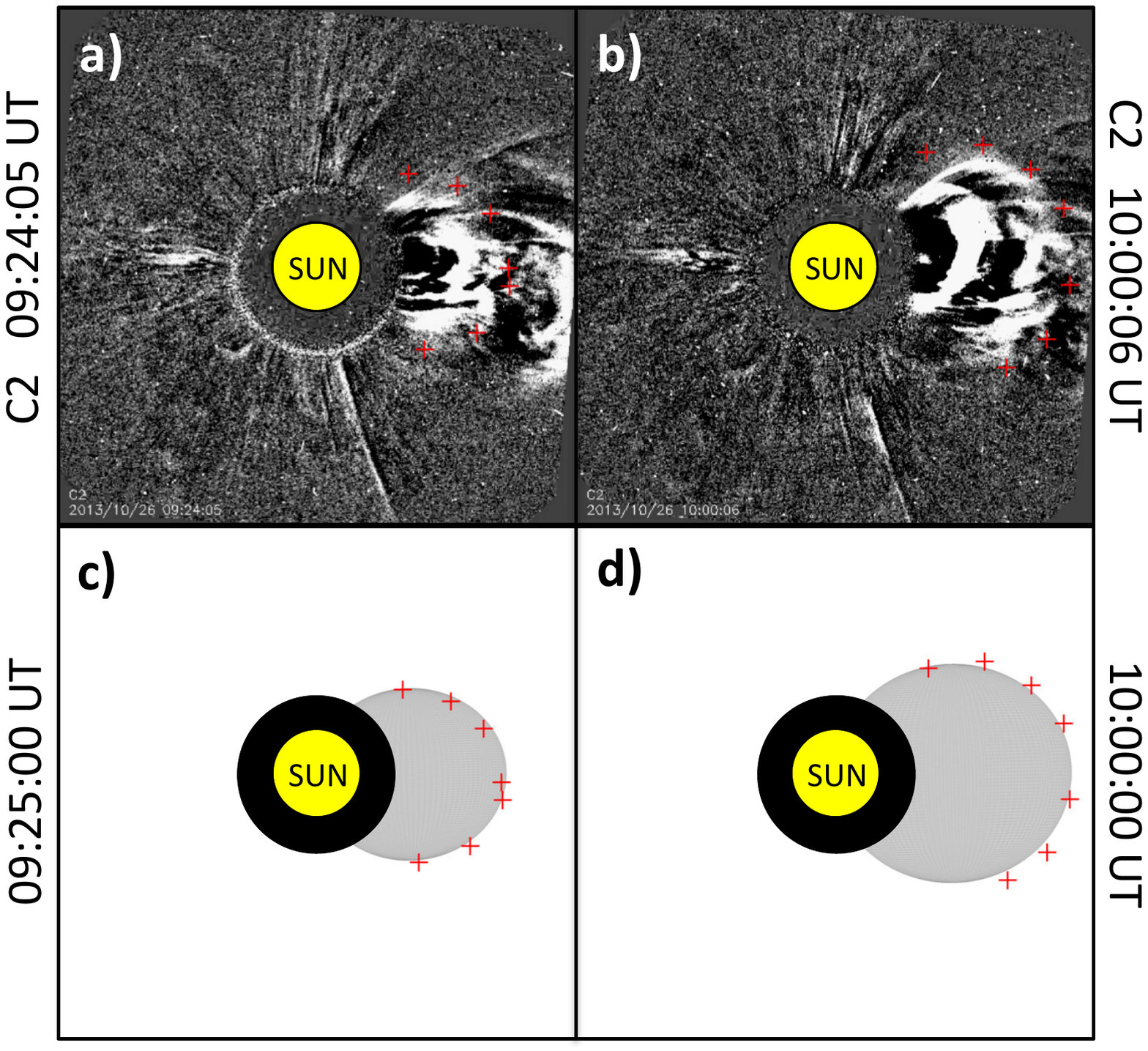}
\caption{(a) Running-difference image of the CME observed with LASCO C2 at 09:24 UT. The superposed red crosses represents the manually selected CME front used to fit the ellipsoid. (b) Same image as panel (b) at 10:00 UT. (c) The resulting projected ellipsoid surface with the related red crosses extracted from the CME running-difference image at 09:25 UT, and at 10:00 UT in panel (d).}
\label{CMEfitting}
\end{center}
\end{figure}

\subsection{CME multi-viewpoint triangulation}
On 2013 October 26, during a period of intense solar activity, a series of CME were observed with the Large Angle and Spectrometric Coronagraph \citep[LASCO;][]{Brueckner1995} on board the \emph{Solar and Heliospheric Observatory} \citep[$SOHO$;][]{Domingo1995} and with the COR2A coronagraph of the SECCHI \citep{Howard2008} instruments on board the STEREO \citep{Kaiser2008} mission. The first CME appeared in LASCO/C2 at 07:00 UT, and it was then followed at 09:12 UT by the CME associated with the type II radio burst presented in this work. A third CME appeared again at 09:48 UT. Owing to the multiple passages of the CMEs, the coronal environment was significantly disturbed. The longitudinal separation between $SOHO$ and $STEREO$-A was 148$\degree$ on 2013 October 26. Based
on this multi-viewpoint dataset, the 3-D surface of the expanding CME can be reconstructed using the method from \citet{Rouillard2016}. Figure~\ref{CMEfitting} shows some examples of the 3D reconstruction technique. The CME front is fitted with an ellipsoid, and panels (a) and (b) show the running-difference images at 09:24 UT and 10:00 UT in which the superposed red crosses were manually obtained to match the CME front observed from LASCO C2. As the CME is propagating in a disturbed corona because of the passage of the previous CME, the white-light front selection was not straightforward and required a careful manual selection. The obtained points that matched the CME front at different times were then used to fit the surface of an ellipsoid at each time step and obtain a set of three parameters. The ellipsoid central position is defined in heliocentric coordinates (radius, latitude, and longitude). These ellipsoids where then visually compared to match the CME observed by the COR2A coronagraph viewpoint. After the parameters of the successive ellipsoids were obtained, we interpolated these parameters at steps of 150 seconds to generate a sequence of regularly time-spaced ellipsoids. To compute the 3D expansion speed of the surface of the CME, we determined for a point P on the ellipsoid at time t the location of the closest point on the ellipsoid at a previous time-step $t-\delta t$ by searching for the shortest distance between point P and all points on the ellipsoid at time $t-\delta t$. We then computed the distance travelled between these two points and divided this by the time interval $\delta t$=150 seconds to obtain an estimate of the speed at P. This approach slightly underestimates the CME front speed, but it returns a simple estimation of the speed in the direction perpendicular to the CME surface.  

\begin{figure}
\begin{center}
\includegraphics[trim=5.5cm 6.5cm 5.5cm 6.5cm,clip=true,width=7.5cm,angle=0]{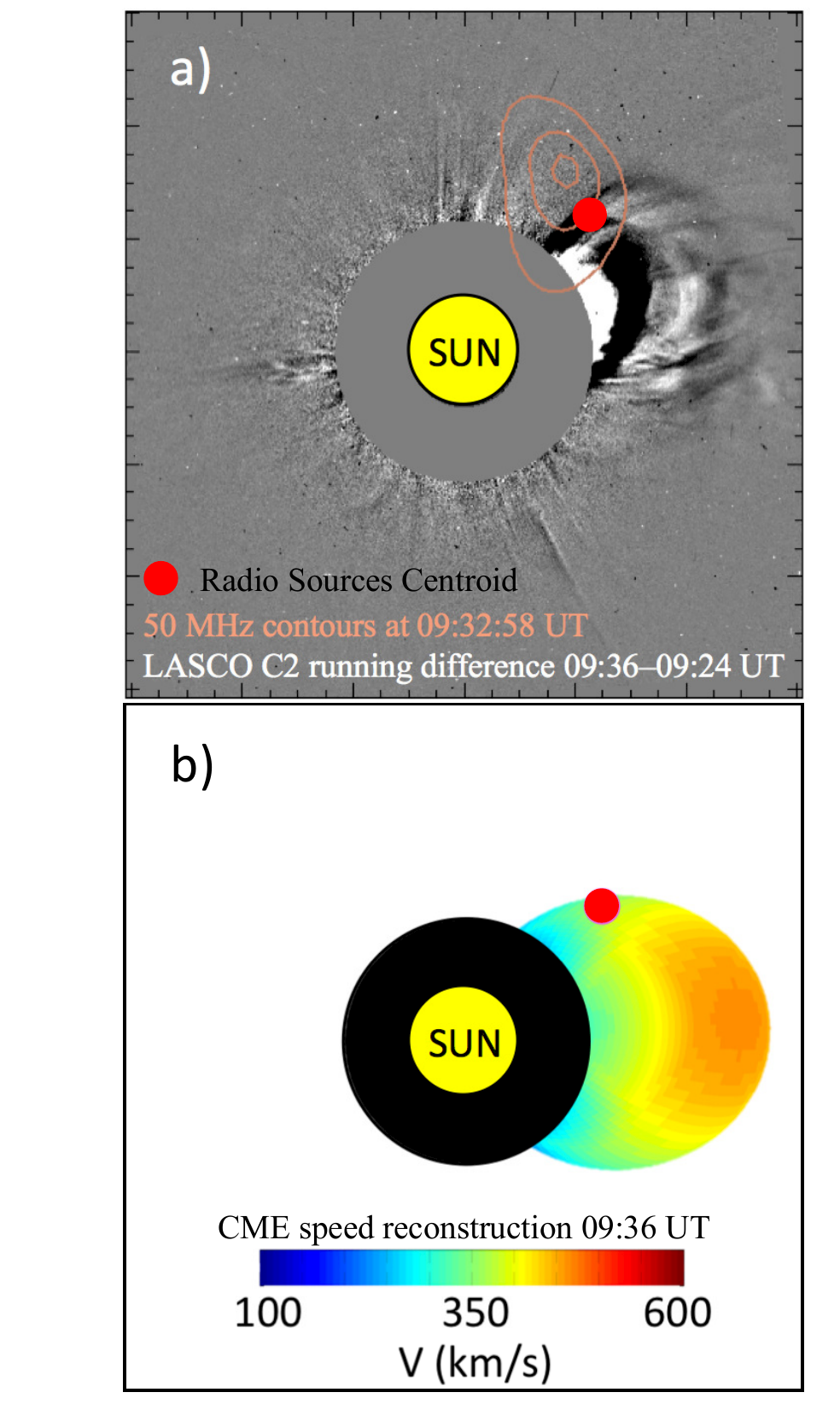}
\caption{(a) {}Running-difference image with superposed contour of the type II harmonic emission at 50 MHz at 09:32:58 UT and the centroid of the position of the radio sources from 70 to 50~MHz from 09:31:10 UT to 09:32:58 (reported in Figure~\ref{spectrum_location}b). (b) CME 3D speed surface reconstruction from multi-viewpoint observations; the radio source centroid is indicated with the pink circle.}
\label{CMEspeed}
\end{center}
\end{figure}

\begin{figure}
\begin{center}
\includegraphics[trim=5cm 2.6cm 4.5cm 1.8cm,clip=true,width=7cm,angle=0]{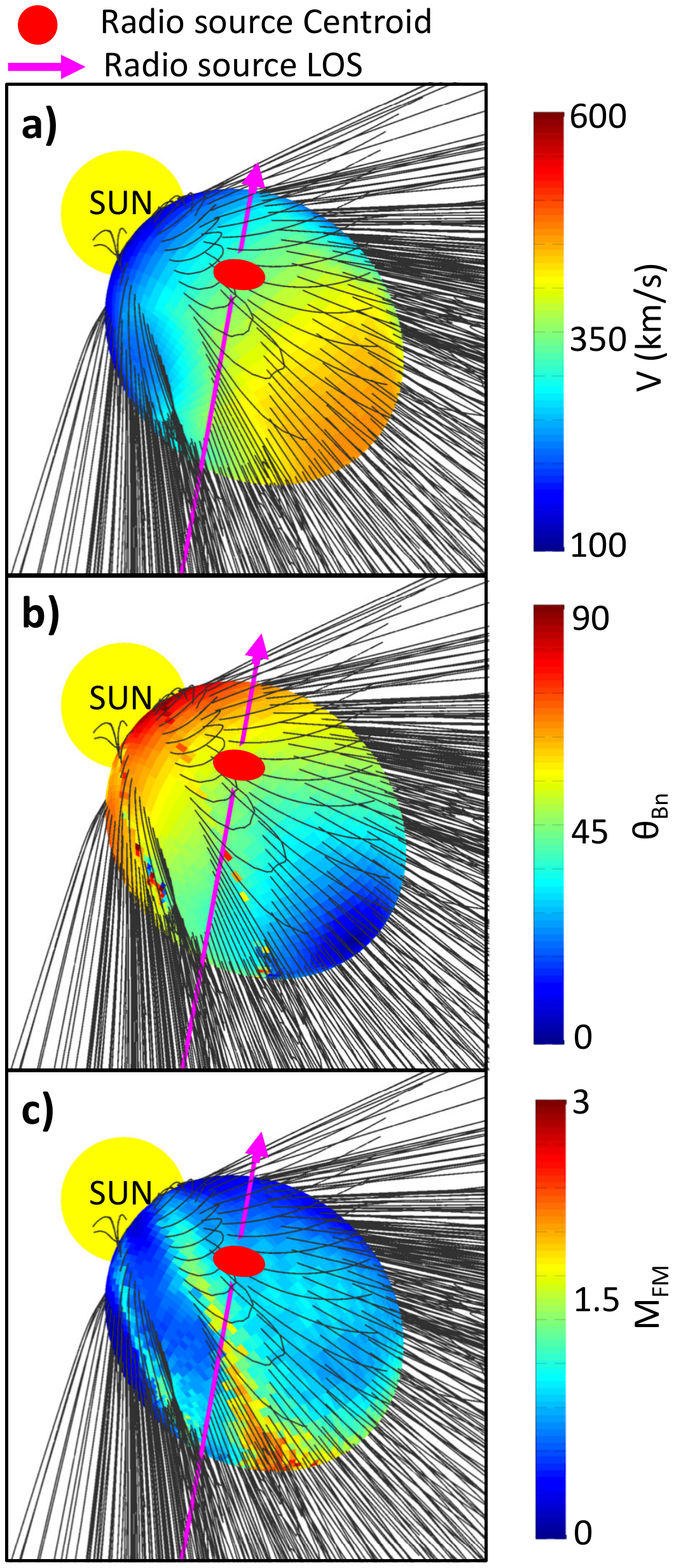}
\caption{3D reconstruction of the coronal ambient parameters. Panel (a) shows the CME speed, panel (b) the magnetic field orientation with respect to the perpendicular direction  to the CME front, and panel (c) shows the fast magnetosonic Mach number. The LOS is indicated with the purple arrow, and the radio source centroid is shown with the red circle. The modelled CME front is reported at 09:36 UT in all panels.}
\label{CME_triangulation}
\end{center}
\end{figure}

Figure~\ref{CMEspeed}a shows the CME running-difference image (09:36 - 09:24 UT) with superposed contours of the 50~MHz source at 09:32:58 UT and the centroid of the radio sources from 70 to 50 MHz between 09:31:10 UT and 09:32:58 UT  (same sources as reported in Figure~\ref{spectrum_location}), indicating the averaged position where the type II emission was recorded (red dot). Figure~\ref{CMEspeed}b compares the location of the type II radio source centroid with the CME expansion speed. A detailed comparison of the radio source positions with the shock front observed in white light is not possible because the time cadence of LASCO/C2 is limited. The type II signature is observed to propagate between 09:31:10 UT and 09:32:58 UT, while the coronagraph images were taken at 9:24 and 9:36 UT. However, even without high time-cadence observations in white light, the radio emission is found to be located at the flank region where the CME expansion speed calculated with the 3D reconstruction is $\sim$370 km s$^{-1}$. The flanks of the CME are not the fastest expanding regions of the CME surface. The apex shows a propagation speed of about 500~km s$^{-1}$. The radio emission, however, is located at the flank of the CME, indicating that other parameters such as the shock geometry and the Mach number play a key role in generating the radio emission from the shock.  To estimate the Mach number along the CME expanding surface, we started reconstructing the ambient corona electron density using a combination of $SDO$/AIA and $SOHO$/LASCO data as described by \citet{Zucca2014}. Subsequently, we used the model called
magneto-hydrodynamic around a sphere polytropic (MASP) developed by Predictive Sciences Inc. \citep{Linker1999}. MASP is a 3 D MHD polytropic model that adopts the photospheric magnetograms from $SDO$/HMI as the inner boundary condition of the magnetic field. The full details of the interpolation technique used to derive the coronal fast-magnetosonic speed from the 3D MHD model results at all points on the surface of the CME is described in \citet{Rouillard2016}.

After we estimated the ambient fast magnetosonic speed, the Mach number was obtained by calculating the ratio between the expanding CME front speed and the fast-magnetosonic speed. Figure~\ref{CME_triangulation} shows a 3D view of the reconstructed expanding CME surface using the technique from \citet{Rouillard2016}. The viewpoint is chosen to allow the overview of the CME speed (a), magnetic field orientation $\theta$ (b), and the fast magnetosonic Mach number (c) of the apex and upper flank of the CME simultaneously. For each panel the line of sight (LOS) is indicated with a purple arrow, and the location of the the type II radio burst is indicated by the red circle. The type II radio emission is recorded at the flank of the CME in a region where the speed of the expanding CME reaches $\sim$370 km s$^{-1}$ and the Mach number ranges from 1.4 to 1.6 and with the orientation of the $B$-field $\theta\sim70\degree$(Figure~\ref{CME_triangulation}).

\begin{figure*}[t]
\begin{center}
\includegraphics[trim=5cm 1cm 5cm 0cm,clip=true,width=7cm,angle=-90]{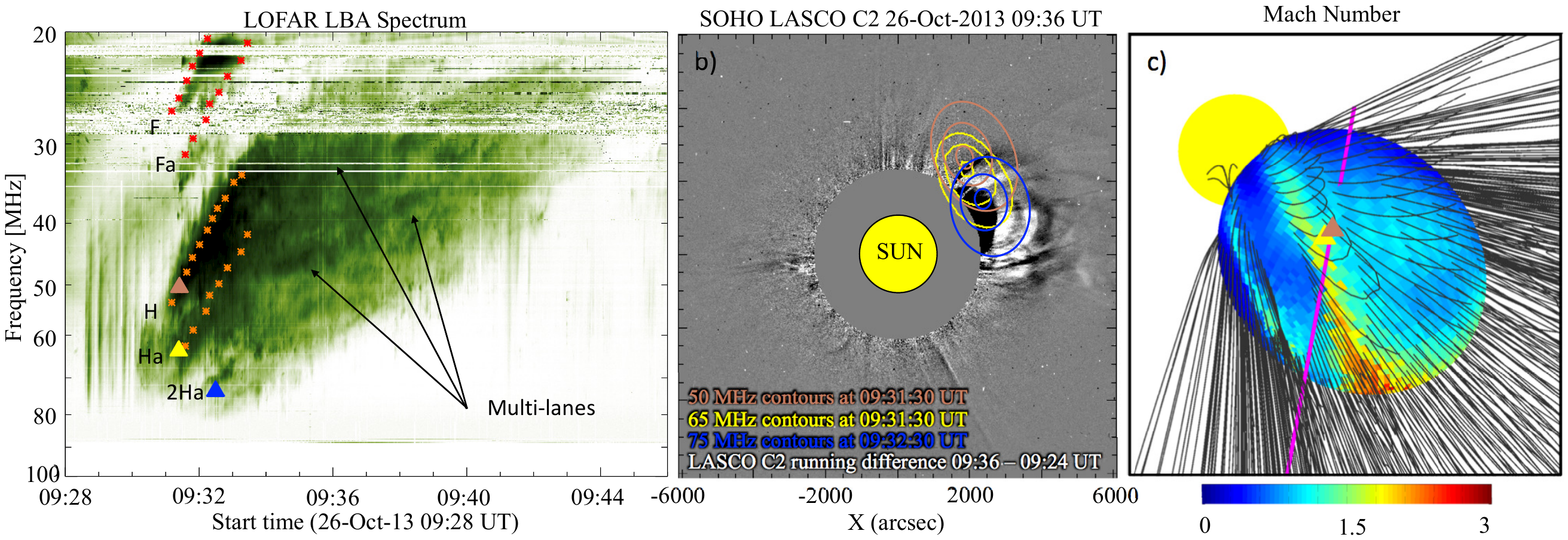}
\caption{(a) Dynamic spectrum of the type II radio burst observed on 2013 Oct 26, showing band-splitting in both the fundamental (F and Fa) and the harmonic lanes (H and Ha) and a series of multi-lanes (indicated by the black arrows). The second harmonic emission can also be seen and is marked with 2Ha. The location of the radio sources for the two split-lanes resolved for 50 and 65~MHz at 09:31:30 UT are indicated in the dynamic spectrum by the brown and yellow triangles, respectively. The second harmonic of the type II radio burst is indicated at 75~MHz with the blue triangle. (b) Running-difference image of the CME observed with $SOHO$/LASCO (09:36 - 09:24 UT) with superposed contours of the radio sources (80\%, 90\%, and 95\%) of the two split lanes H and Ha at 09:31:30 UT. The source location of the second harmonic at 09:32:45 using the colour code of the triangles in panel (a) is also shown. (c) Reconstructed CME Mach number. The locations of the type II split-lanes are superposed. The LOS is indicated by the purple line.}
\label{spectrum_bandsplitting}
\end{center}
\end{figure*}

\subsection{Band-splitting and multi-lanes}
Type II radio bursts typically present two bright bands of emission with a frequency ratio of $\sim$2. These are commonly accepted to be the emission of the fundamental and first harmonic of the local plasma frequency \citep[see e.g.][]{Wild1972,Mann1995,Aurass1997,Gopalswamy2000,Pick2008}. These two bright bands often appear to show a distinct separation into two sub-bands with an average frequency ratio of $\sim$1.23 \citep{Vrsnak2001,Du2015} in both fundamental and harmonic emission. This phenomenon is known as band-splitting of the type II radio burst, and its origin is still controversial. One explanation that found observational evidence in the past \citep{Smerd1974,Vrsnak2001} and recently \citep{Zimovets2012,Zucca2014b} is that the two lanes originate from a simultaneous radio emission from the upstream and downstream region of a shock. \citet{Holman1983} suggested that band-splitting might also result from a planar shock front moving non-radially across curved magnetic field lines. Another interpretation of band-splitting is that the two sub-lanes are the result of emission from two different parts of the shock front with similar expanding speeds, where the coronal ambient properties such as the electron density, magnetic field, and Alfv\'en speed are different \citep[see e.g.][]{McLean1967,Schmidt2012}. 
In addition to the band-splitting phenomenon, type II radio bursts may also show multiple separate lanes \citep{Nelson1985}. These multiple lanes cannot be explained with a simple upstream and downstream emission from a single shock front. Recently, \citet{Zimovets2015} reported an observation with three separate lanes. Using observations from the Nan\c cay radioheliograph \citep{Kerdraon1997}, they showed distinct locations of the radio sources associated with the different lanes. They proposed a scenario in which two lanes are paired and originate from the upstream and downstream of a shock front, while the third lane was found to have its origin in another location and resulted from a different shock front.

The type II burst observed on 2013 October 13 with LOFAR presents both band-splitting and multiple lanes. A dynamic spectrum showing the band-splitting and multiple lanes in the type II burst is shown in Figure~\ref{spectrum_bandsplitting}a. Band-splitting is visible in both fundamental (indicated with red stars and marked F and Fa) and harmonic emission (indicated with orange stars and marked H and Ha). The type II radio burst shows also a second harmonic band of emission, marked 2Ha. This is the second harmonic emission of the Fa lane, while the second harmonic emission of the F lane is not clearly discernible as it is superposed with the Ha emission lane. Multiple lanes other than the different orders of emission of the F and Fa pair are also present. These are indicated in the dynamic spectrum with the black arrows, and they are evident starting from 09:34 UT, where the type II signature becomes more complex. 
The radio source positions for the lanes H and Ha and the second harmonic 2Ha are indicated in panel (b) using the same colour code as the triangles in panel (a), while no spatial information could be inferred for the F and Fa bands with the current beam setting as these lanes are below 50~MHz. Sources are superposed over the LASCO white-light running-difference image of the CME at 09:36 UT in panel (b). Lanes H and Ha (the temporal evolution of lane Ha is reported in Figure~\ref{spectrum_location}) resulting from band-splitting are located in the flank of the CME. Panel (c) shows the fast magnetosonic Mach number estimated with the 3D reconstruction over the CME surface. The positions of the radio sources of the band-splitting lanes (H and Ha) are superposed in the 3D reconstruction with the two triangles, using the colour code of the triangles in panel (a) and indicating the LOS path with a purple line. The two source locations are in a region where the Mach number is higher than 1. In particular, lanes H and Ha are located in a region where the fast magneto-sonic Mach number is in the range 1.4--1.6. Within the resolution of this observation, the two sources (H and Ha) are located in the same region. This is in agreement with the band-splitting interpretation of emission ahead and behind the shock front. When we consider this interpretation, the fast magnetosonic Mach number can be estimated by calculating the compression ratio inferred from the frequency split in the type II harmonic lane. The average value of the compression ratio from the band-splitting along the first harmonic emission lanes is $X$=1.45. We estimated the fast magneto-sonic Mach number using the method described by \citet{Vrsnak2002}, using equation (9) of their manuscript. The fast magnetosonic Mach number $M_{FM}=M_A/(1+\beta\gamma/2)^{1/2}$ assuming a polytropic index $\gamma=5/3$, has a value of 1.32--1.35, using a plasma-to-magnetic pressure ratio $\beta$ between 0 and 2, respectively. For a value of $\gamma=1$ (uniform coronal temperature), the Mach number is estimated as $M_{FM}$=1.36--1.54 and is in this case comparable with the Mach number estimated with the 3D CME reconstruction (1.4--1.6). 

\subsection{Second harmonic emission}
The second harmonic emission (three times the fundamental emission) is rarely observed in type II radio bursts. The type II radio burst observed on 2013 October 26 by LOFAR showed this emission in the dynamic spectrum. This emission lane is marked as 2Ha in Figure~\ref{spectrum_bandsplitting}. The tied-array beam source reconstruction was used to locate the source position of the second harmonic lane 2Ha. The radio source position at 75~MHz of the 2Ha lane is indicated with blue contours in Figure~\ref{spectrum_bandsplitting}b. The position is comparable within the beam size with the location of the first harmonic lane Ha, suggesting that the second and third harmonic emission originate from the same source region. This finding is in agreement with \citet{Zlotnik1998}, where the authors were able to measure the source location of another type II burst showing first and second harmonic emission. Using the NRH imaging bands at 327~MHz and 435~MHz, they concluded that their observations were in agreement with the first and second harmonic lane of the type II radio burst coming from the same source. This confirmation using LOFAR suggests that radio wave propagation does not significantly alter the apparent location of the harmonic source even at low frequencies such as 65 and 75~MHz.

\section{Conclusion}
\label{discussion}
We have presented a study of the type II radio burst observed on 2013 October 26. We were able for the first time to estimate the spatial location of the type II radio burst at frequencies between (70--50 MHz) using the LOFAR LBA antennas. This complex type II radio burst was composed of a pair of split lanes observed at fundamental, first, and second harmonic emission and of several multi-lanes. The location of the type II radio signature was compared with the CME front reconstructed in 3D. The fast magnetosonic speed and the $B$-field orientation were used to estimate the shock Mach number and the shock magnetic field geometry along the CME surface. We have found that the radio signature of the shock is located at the flank of the CME. In particular, the band-splitting of the type II first harmonic emission is located in the flank of the CME where the Mach number ranges between 1.4 and 1.6 and the configuration is quasi-perpendicular, $\theta_{Bn} \sim$70 degrees. 
This study provides observational evidence on the location of the type II emission in the region with quasi-perpendicular geometry and Mach number greater than 1. This is the region where particles can efficiently be accelerated to higher energies by shock drift acceleration \citep{Holman1983,Mann2005}. The quasi-perpendicular geometry related to a signature of a type II radio burst was recently found also by \citet{Salas2016}. In addition, multiple lanes in the type II emission may be explained with the radio signature coming from different regions of the shock front where the local plasma conditions are favourable for quasi-perpendicular shocks and for the generation of associated radio emission. However, the observations we presented do not allow identifying the spatial position of these lanes. We were able to locate the source position of the first harmonic split lanes, which was found to be consistent with the emission ahead and behind the expanding shock front in the CME flank. A comparable Mach number (1.3--1.5) was calculated independently from the band-splitting in the dynamic spectrum assuming this scenario. However, other scenarios cannot be excluded since the observation we presented does not provide a definite answer to the band-splitting phenomenon. The location of the second harmonic emission was also identified. It was located within the beam size in a common source region with the first harmonic emission. This confirms previous findings and excludes that in this event, radio wave propagation significantly alters the apparent location of the radio source of the second harmonic emission. Radio observations with better resolution are necessary to clearly describe the origin of the different emission lanes and to interpret them. In particular, LOFAR observations using simultaneously imaging and tied-array beam will reduce the uncertainties of the radio source location, and will allow determining the location of fainter radio sources. Questions remain about the number of events with shocks at the flank of CMEs, about the necessity of a quasi-perpendicular geometry in type II emission, and about the nature of the band-splitting and multi-lane phenomena. Multi-viewpoint observations together with imaging campaigns using LOFAR are important for solving the remaining unknowns of the type II radio emission, the related fine structures, and the relationship between CME expansion and ambient medium parameters in producing the radio emission.

\begin{acknowledgements}
This paper is based on data obtained with the International LOFAR Telescope (ILT). LOFAR (van Haarlem et al. 2013) is the Low Frequency Array designed and constructed by ASTRON. It has facilities in several countries that are owned by various parties (each with their own funding sources), and that are collectively operated by the ILT foundation under a joint scientific policy. We are also grateful to the STEREO, SDO, and LASCO team for the data access. A. P. Rouillard acknowledges funding from the French ``Agence Nationale de la Recherche" under contract number ANR-17-CE31-0006-01. 

\end{acknowledgements}

\email{zucca@astron.nl}.

\bibliographystyle{aa}
\bibliography{mybib.bib} 
\clearpage

\end{document}